\documentclass[prb, 11pt]{revtex4}
\usepackage{graphicx}
\usepackage{dcolumn}
\usepackage{bm}
\begin{document}
\title{A first-principles investigation of the thermodynamic and mechanical properties of Ni-Ti-Sn Heusler and half-Heusler materials}
\author{P. Hermet, K. Niedziolka and P. Jund}
\affiliation{Institut Charles Gerhardt Montpellier, UMR 5253 CNRS-UM2-ENSCM-UM1, Universit\'e Montpellier 2, Place E. Bataillon, 34095 Montpellier C\'edex 5, France}
\begin{abstract}
  First principles calculations of the vibrational, thermodynamic and
  mechanical properties of the Ni-Ti-Sn Heusler and half-Heusler
  compounds have been performed. First, we have calculated the Raman
  and infrared spectra of NiTiSn, providing benchmark theoretical data
  directly useful for the assignments of its experimental spectra and
  clarifying the debate reported in the literature on the assignment of its
  modes. Then, we have discussed the significant vibrational 
  density-of-states of Ni$_2$TiSn at low-frequencies. These states are at the origin
  of (i) its smaller free energy, (ii) its higher entropy, and (iii)
  its lower Debye temperature, with respect to NiTiSn. Finally, we
  have reported the mechanical properties of the two compounds. In
  particular, we have found that the half-Heusler compound has the
  largest stiffness. Paradoxically, its bulk modulus
  is also the smallest. This unusual behavior has been related to the 
  Ni-vacancies that weaken the structure under isostatic compression. Both compounds show a ductile behavior.
\end{abstract} \maketitle
\section{Introduction}
\begin{figure}
\begin{center}
\includegraphics[width=16cm]{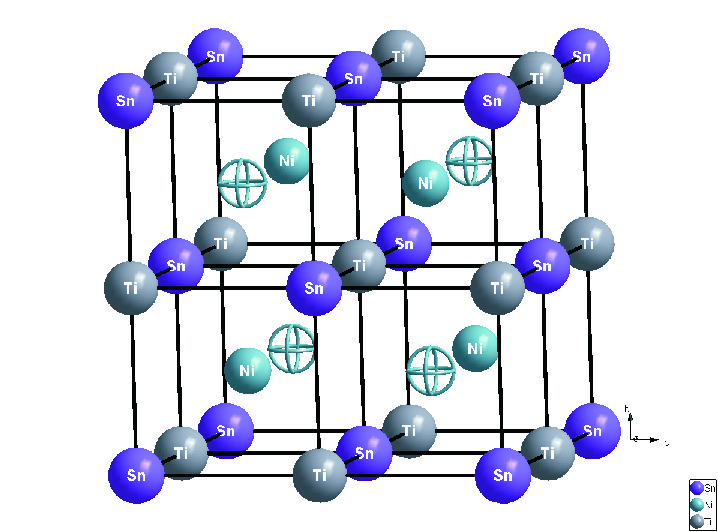}
\caption{Structure of NiTiSn and Ni$_2$TiSn in which the open spheres are occupied by Ni atoms. Color of Ni, Ti and Sn-atoms is cyan, gray and purple, respectively.}
\label{structure}
\end{center}
\end{figure}

The so-called Heusler and half-Heusler compounds have attracted
intensive work during the last years. These compounds are a class of
ternary intermetallics associating three elements in the following
stoichiometric proportions 1:1:1 (Half-Heusler) or 2:1:1
(full-Heusler). They are represented by the general formula: XYZ and
X$_2$YZ, where X and Y are transition elements, and Z is a $s$ or
$p$-element. From the structural point of view, Heusler
(resp. half-Heusler) compounds generally crystallise in the Fm$\bar3$m
(resp. F$\bar 4$3m) space group with Cu$_2$MnAl (resp. MgAgAs) as
prototype~ \cite{pearson} (see Fig.~\ref{structure}).

New properties and potential fields of application constantly emerge
\cite{properties} in these materials: topological insulators and
spintronics are recent examples. Their properties can be easily
predicted by the valence electron concentration
(VEC)~\cite{Felser} and their extremely flexible electronic
structure offers a lot of possibilities for tailoring these materials
for interesting physical applications \cite{properties}(a detailed analysis of the density of states 
has been done in a separate 
study~\cite{Colinet}). Concerning
more specifically the thermoelectric properties which we are
interested in, it has been demonstrated that a value of eighteen for the
valence electron concentration leads to potentially good
materials~\cite{Xia}. Since good thermoelectric materials are
typically heavily doped semiconductors (such as degenerated
semiconductors), the classes of materials which are presently under
investigation include mainly half-Heusler compounds. Half-Heusler
compounds, like MNiSn (M = Ti, Zr, Hf), are $n$-type semiconductors
with a narrow bandgap (Eg= 0.1--0.5~eV), a high Seebeck
coefficient ($\approx$ -200$\mu$V/K) and a low electrical resistivity
($\approx$ 10$^{-4}$ $\Omega$m). Unfortunately at this time, a
relatively high thermal conductivity is known to be around 10 W/mK at
300~K~\cite{properties,thermo}.  Recently, Wee {\it et al.}~\cite{Wee}
have estimated the thermal conductivity of NiTiSn from a semi-analytic
model and have confirmed the experimental findings. A drastic decrease
of the thermal conductivity associated to good electronic transport
properties is therefore mandatory if this material is to be used in
thermoelectric applications \cite{thermo2}.

An important contribution to the thermal conductivity is the lattice
contribution which is directly connected to the vibrational properties
of the compound. In this context, we report, in this paper, a complete 
theoretical study of the lattice dynamics of Ni$_2$TiSn and NiTiSn, as
well as of their thermodynamic and mechanical properties. Our aim is to provide
benchmark theoretical data, directly useful for further studies on
materials belonging to the Heusler and half-Heusler classes. Surprisingly,
no comparison between these two compounds has been proposed so
far. Indeed the few studies reported in the literature are devoted to
investigate the phonon modes of NiTiSn~\cite{Popovic,Mestres,Wee},
whereas those of Ni$_2$TiSn remain presently unexplored to our
knowledge, even at the zone-center. Thus, we have studied the
correlation between the zone-center phonons and the crystal structures
of Ni$_2$TiSn and NiTiSn. For the latter, the calculations of its
Raman and infrared spectra allowed us to clarify the
debate reported in the literature on the assignment of its phonon modes. Then,
within the quasi-harmonic approximation, we have studied the thermodynamic
properties of Ni$_2$TiSn and NiTiSn, and we bring new insights into
the phonon modes at the origin of their differences. Finally, their
mechanical properties are investigated as well. We
show that the response of a given material depends on the nature of
the stress, which is not only interesting from a fundamental point of
view but also has some consequences when the materials are to be used
in practical applications.
 
\section{Computational details }

First-principles calculations of Ni$_2$TiSn (metallic compound) and
NiTiSn (semiconducting compound) were performed within the density
functional theory (DFT) framework as implemented in the ABINIT
package~\cite{ABINIT}. The exchange-correlation energy functional was
evaluated using the generalized gradient approximation (GGA)
parametrized by Perdew, Burke and Ernzerhof (PBE)~\cite{PBE} or the
local density approximation (LDA) parametrized by
Perdew and Wang~\cite{PW}. The all-electron potentials were replaced by
norm-conserving pseudopotentials generated according to the
Troullier-Martins scheme~\cite{TM}. Ni($3d^{8}$, $4s^2$), Ti($3d^{2}$,
$4s^2$), and Sn($5s^{2}$, $5p^2$)-electrons were considered as valence
states. The electronic wave functions were expanded in plane-waves up
to a kinetic energy cutoff of 65~Ha and integrals over the Brillouin
zone were approximated by sums over a 8$\times$8$\times$8 mesh of
special $k$-points according to the Monkhorst-Pack
scheme~\cite{Monkhorst}. A Fermi-Dirac scheme with a
smearing width equal to 0.01~Ha was used for the metallic occupation
of Ni$_2$TiSn.

Dynamical matrix, dielectric constants, Born effective charges and
elastic tensors were computed within a variational approach to density
functional perturbation theory~\cite{Gonze97}. Phonon dispersion
curves were interpolated according to the scheme described by Gonze
{\it et al.} \cite{Gonze}. In this scheme, the dipole-dipole interactions are subtracted from the dynamical matrices before Fourier transformation, so that only the short-range part is handled in real space. A 4 $\times$ 4 $\times$ 4 $q$-points grid in the irreducible Brillouin zone was employed for the calculation of the vibrational band structure and the phonon density-of-states (DOS),
whereas a denser 120 $\times$ 120 $\times$ 120 $q$-points grid was
used for the evaluation of the integrals associated to the
thermodynamical properties. The intensity of the NiTiSn Raman
lines~\cite{Veithen04,Hermet06,BTO} was obtained at the
LDA level~\cite{Note} and within a nonlinear response formalism taking advantage
of the 2$n$+1 theorem.

Structural relaxations were performed until the maximum stresses were
less than 1$\times$10$^{-4}$~GPa. Our relaxed LDA lattice parameters
($a=$5.91~\AA\ for Ni$_2$TiSn and $a=$5.72~\AA\ for NiTiSn)
underestimate the experimental ones of about -3\%
($a_{exp}=$6.09~\AA\ for Ni$_2$TiSn~\cite{gorlich} and $a_{exp}=$5.92~\AA\
for NiTiSn~\cite{ouardi}), whereas GGA ($a=$6.21~\AA\ for Ni$_2$TiSn
and $a=$6.00~\AA\ for NiTiSn) overcorrects the LDA predictions leading
to lattice parameters about +2\% larger. These trends are usual in DFT.  Our calculated electronic
band gap of NiTiSn at the GGA level (E$_g^{GGA} = $0.49~eV) is strongly overestimated with
respect to the experimental value (E$_g^{exp} \approx $0.12~eV) reported by
Aliev {\it et al.}~\cite{aliev} . This unusual behaviour is not related
to a convergence problem of our calculations, but is rather a
consequence of the existence of structural defects and/or impurities in NiTiSn crystals
that are not considered in our calculations. Our overestimation of the
experimental band gap is consistent with previous first-principles
calculations of the electronic structure of
NiTiSn~\cite{ouardi,gap} using the GGA exchange--correlation.
 
\section{Results and discussion}

\subsection{Zone-center optical phonon modes}

\begin{figure}
\begin{center}
\includegraphics[width=16cm]{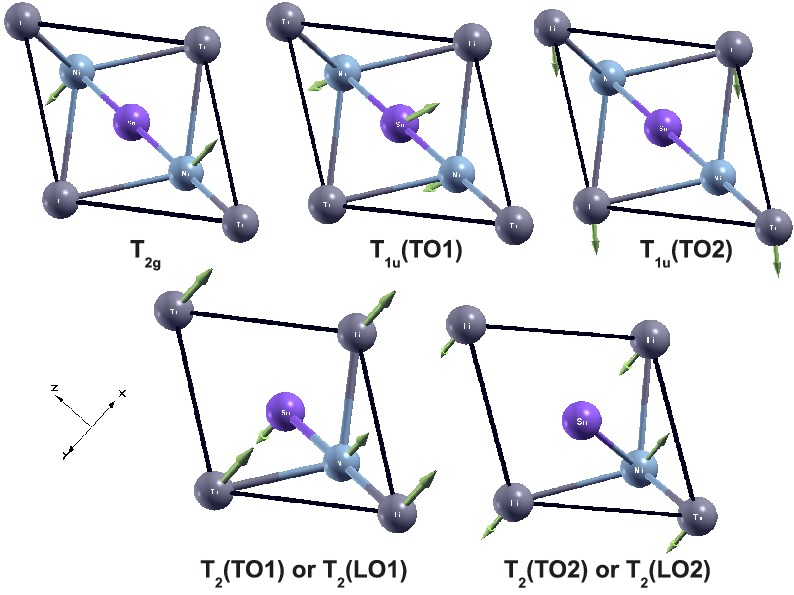}
\caption{Normal modes of Ni$_2$TiSn (top) and NiTiSn (bottom). Arrows are proportional to the amplitude of the atomic motions. Color of Ni, Ti and Sn-atoms is cyan, gray and purple, respectively.}
\label{anim}
\end{center}
\end{figure}

At the zone-center, optical phonon modes of Ni$_2$TiSn can be
classified, according to the irreducible representations of the O$_h$
point group, into: $\Gamma_{opt} = T_{2g} \oplus 2 T_{1u}$. These
modes are triply degenerated and they are either infrared (T$_{1u}$)
or Raman (T$_{2g}$) active. The Raman mode is calculated at
122~cm$^{-1}$ and it is assigned to antisymmetric motions of
Ni-atoms (see Fig. \ref{anim}). The frequencies of the two infrared modes are calculated at
181 and 237~cm$^{-1}$. The first mode involves symmetric motions of
Ni-atoms counterbalanced by motions of Sn-atoms, whereas the second
mode is only dominated by motions of Ti-atoms (Fig. \ref{anim}). However these frequencies cannot 
be compared to experiment or previous calculations because
Raman or infrared spectra of Ni$_2$TiSn are presently unavailable in
the literature. No further investigation of its zone-center phonon
modes, like the calculation of infrared and Raman intensities, can be
performed within our formalism (Berry phase) due to the metallic
character of this compound~\cite{Note}.

In contrast to Ni$_2$TiSn, the NiTiSn structure is represented by four
interpenetrating cubic FCC sublatices, one of which being occupied
by Ni-vacancies (see Fig.~\ref{structure}). This structure is therefore noncentrosymmetric and
belongs to the T$_d$ point group. In this case, the previous T$_{2g}$ mode
is no longer possible, and the compatibility relations between the
O$_h$ and T$_d$ groups lead to: T$_{1u}$ $\rightarrow$ T$_2$. Thus,
the irreducible representation of the zone-center optical phonons of
NiTiSn is reduced to: $\Gamma_{opt} = 2 T_2$, where each of these two
triply degenerate modes can be both infrared and Raman active. In the
literature, phonon assignments of NiTiSn were performed by Raman and
infrared spectroscopies using a filiation procedure between three
compounds: MeNiSn, with Me = Ti, Zr, Hf~\cite{Popovic,Mestres}. Accordingly, the frequency of the lowest longitudinal optical (LO)
phonon mode is not yet unambiguously identified and some controversies
remain about the assignment of the two transverse optical (TO) modes
and the remaining LO mode. Thus, to have a reliable assignment of the NiTiSn modes, we performed calculations at LDA and
GGA levels to estimate the dependence of the  phonon
frequencies on the standard
exchange--correlation (XC) functionals. These results are listed in Table \ref{freq} with the
experimental data and a previous first-principles calculation reported
in the literature~\cite{Wee}. We observe that most of our frequencies
are systematically downshifted with respect to the calculation from
Wee {\it et al.}~\cite{Wee}, but the two sets of results are however
consistent. We attribute these frequency shifts to the
different equilibrium structures obtained using different convergence
parameters and pseudopotentials~\cite{PatBFO}. We observe that the
experimental TO modes are reasonably well predicted by the different
XC-functional classes. The experimental frequencies are bounded
between those predicted by LDA and GGA with relative errors smaller
than 8\%. A slightly better agreement with the experimental
frequencies is however obtained at the GGA level. The TO1 mode
involves motions of the three kinds of atoms, and we assign it to a
TiNiSn scissoring (see Fig. \ref{anim}). This assignment is consistent with
the Ti-Ni-Sn motions reported by Popovic {\it et al.}~\cite{Popovic}
and Wee {\it et al.}~\cite{Wee}. We assign the TO2 mode to a TiNiTi out-of-phase wagging where Sn-atoms
are not involved (see Fig. \ref{anim}). This assignment is in agreement with
Wee {\it et al.}~\cite{Wee}, but in disagreement with Popovic {\it et al.}~\cite{Popovic} who reported opposite motions of Ti and Sn-atoms.

\begin{figure}
\begin{center}
\includegraphics[width=16cm]{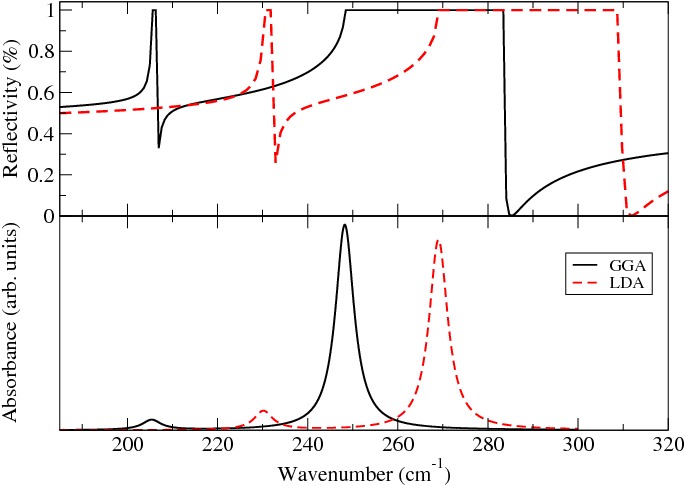}
\caption{Reflectivity (top) and absorption (bottom) infrared spectra of NiTiSn.}
\label{IR}
\end{center}
\end{figure}

Frequencies of the calculated LO-modes are also reported in Table \ref{freq}
with the experimental values obtained from the fit of the infrared
reflectivity spectrum. Their prediction is expected to be less
accurate than for the TO-modes due to typical LDA/GGA
inaccuracies. Indeed, the LO-TO splitting depends on the knowledge of
the electronic dielectric tensor and on the Born effective charges
observed to be dependent on the used
functional~\cite{PatBFO}. However, it is interesting to note that both
LDA and GGA lead to the same frequency of the LO2 mode
(284~cm$^{-1}$), which is close to the experimental values reported in
the 285--292~cm$^{-1}$ range.  Reflectivity spectrum is calculated at
normal incidence according to the methodology from Ref. \cite{PatNiOH}
and is displayed in Fig. \ref{IR}. Since our approach neglects the
damping of the phonon modes, the calculated reflectivities saturate to
unity. We find a relative good agreement between our calculation and
the experimental spectra reported in the
literature~\cite{Mestres,Popovic} (not shown here). We clearly observe
that the LO-TO splitting of the TO1-mode is negligible (around
1~cm$^{-1}$), whereas it reaches 36~cm$^{-1}$ for the TO2-mode at the
GGA level. These observations are in agreement with experimental
data~\cite{Mestres,Popovic} and the calculation from Wee {\it et
  al.}~\cite{Wee}. Concerning now the assignment of the LO-modes to
specific motions, we can convincingly assign the LO1 and TO1 modes to
the same atomic motions due to their negligible splitting (see Fig.
\ref{anim}). However, in the case of the LO2-mode, its
eigendisplacement vectors do not necessarily correspond to the ones of its
corresponding TO2-mode due to long-range Coulomb
interactions. The possible mixing between the LO2 mode and the TO
modes has been calculated according to the overlap
matrix~\cite{Hermet06}:
\begin{equation}
\langle\mathbf u_{LO2} | M | \mathbf u_{TOn} \rangle = \sum_{\alpha,\kappa} u_{LO2}(\kappa\alpha) M_\kappa u_{TOn}(\kappa\alpha),
\end{equation}
where the sum runs over the space directions $\alpha$ and the atoms
$\kappa$, $M_\kappa$ is the mass of the $\kappa^{th}$ atom, and
$\mathbf u_{LO2}$ (resp. $\mathbf u_{TOn}$, where $n= $1, 2 ) are the
eigendisplacement vectors of the LO2-mode (resp. the two TO modes). We found
that this mixing is, however, negligible as the LO2-mode overlaps at 99\% with
the TO2 mode. Thus, as in the case of the LO1-mode, we conclude that
the LO2 and the TO2 modes have the same assignment (see Fig. \ref{anim}).

\begin{figure}
\begin{center}
\includegraphics[width=16cm]{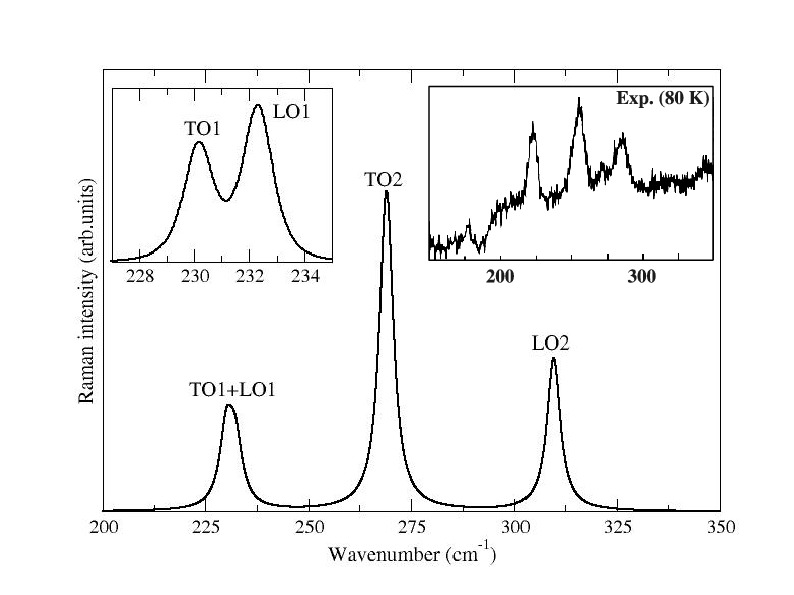}
\caption{Calculated LDA Raman spectrum of NiTiSn using a Lorentzian line shape and a constant linewidth fixed at 2~cm$^{-ˆ1}$. Inset: (right) Experimental spectrum  recorded by Mestres {\it et al.}~\cite{Mestres} at 80~K, (left) Calculated LDA Raman spectrum in the 227--235~cm$^{-1}$ range using a smaller linewidth fixed at 1~cm$^{-ˆ1}$.}
\label{Raman}
\end{center}
\end{figure}

Fig. \ref{Raman} compares the calculated and experimental Raman spectra of
NiTiSn. Frequency positions and relative intensity of the experimental
lines are well reproduced by our calculation. The experimental
spectrum is dominated by three strong lines centered at 220, 255 and
285~cm$^{-1}$. The two first lines are assigned to TO-modes by Popovic
{\it et al.}~\cite{Popovic} and Mestres {\it et al.}~\cite{Mestres},
while only the last authors have observed the line at 285~cm$^{-1}$
assigned to a LO-mode. Our calculations unambiguously support the
identification of the TO2 and LO2 lines. In the case of the LO1 line,
our calculations show that its intrinsic scattering efficiency should
be similar to that of the TO1 line (see Table \ref{sus}). Thus, the LO1 line
remains experimentally unobserved because it overlaps with the TO1
line due to their very close frequencies ($\Delta \omega \approx$ 1~cm$^{-1}$). This
overlap can be avoided for the calculated spectrum since we use a
constant linewidth to represent the Raman lines (see inset of
Fig. \ref{Raman}, left). The first experimental line at 220~cm$^{-1}$
should be therefore identified as a juxtaposition of the TO1 and LO1
modes. Five weak lines, not predicted by the group theory (and so not obtained in
our calculations), can also be observed in the experimental spectrum
around 180, 230, 275, 320 and 340 cm$^{-1}$. They could therefore be
associated to structural defects, inhomogeneities, or impurities in
the NiTiSn compound.

To conclude this section, we analyzed the static dielectric
permittivity, $\varepsilon^0$, of NiTiSn. This tensor can be decomposed as the sum of an
electronic ($\varepsilon^{\infty}$) contribution and a contribution of
each individual phonon mode ($\varepsilon_m^{ph}$) such
as~\cite{Gonze97,PatNiOH}:
\begin{equation}
 \varepsilon_{\alpha\beta}^0=\varepsilon_{\alpha\beta}^{\infty}+\sum_m \varepsilon_{\alpha\beta,m}^{ph} = \varepsilon_{\alpha\beta}^{\infty}+\frac{4\pi}{\Omega_0}\sum_m \frac{S_{\alpha\beta}(m)}{\omega_m^2},
\end{equation}
where the sum runs over all modes $m$, $\Omega_0$ is the unit cell
volume and $S$ is the infrared oscillator strength. Results of this
decomposition are reported in Table \ref{FOS}. We observe that
$\varepsilon^0$ is mainly governed by the electronic contribution
($\approx$ 75\%). The latter is significantly underestimated at the LDA and GGA levels ($\varepsilon^{\infty}_{calc}
\approx $21) with respect to the experimental value
($\varepsilon^{\infty}_{exp} = $36.5) reported by Popovic {\it et
  al.}~\cite{Popovic} and obtained from the fit of the infrared
reflectivity spectrum. This underestimation is consistent with our
overestimation of the electronic band gap of NiTiSn, and this problem has been
discussed in Sec. II.
The phonon contribution to $\varepsilon^0$ is mainly dominated by
the TO2 mode (93\%) which combines the largest mode effective charge
and oscillator strength. This mode therefore dominates the infrared
absorption spectrum of NiTiSn, as observed in Fig. \ref{IR}.

\subsection{Phonon density-of-states and thermodynamic properties}

\begin{figure}
\begin{center}
\includegraphics[width=16cm]{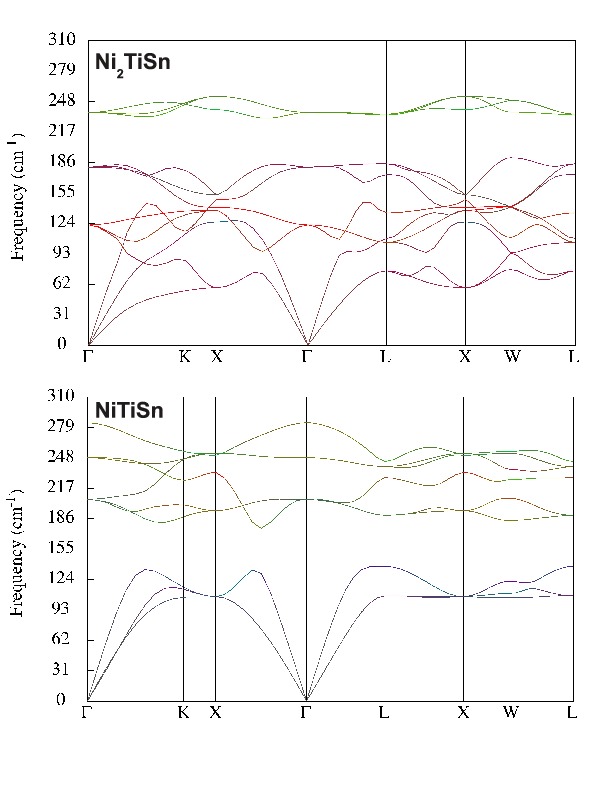}
\caption{Phonon dispersion curves of Ni$_2$TiSn and NiTiSn calculated at the GGA level. A color has been assigned to each point based on the contribution of each kind of atom to the associated dynamical matrix eigenvector: red for the Ni-atoms, green for the Ti-atoms, and blue for Sn-atoms.}
\label{VBS}
\end{center}
\end{figure}

\begin{figure}
\begin{center}
\includegraphics[width=16cm]{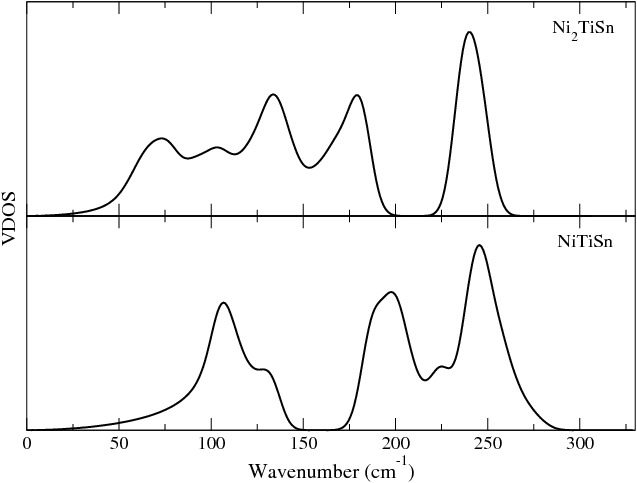}
\caption{Calculated GGA phonon density-of-states of Ni$_2$TiSn and NiTiSn.}
\label{VDOS}
\end{center}
\end{figure}

In this section, only our results obtained at the GGA level will be discussed, as they are in better
agreement than the LDA ones with the experimental data (lattice parameters
and zone-center TO phonons). Phonon dispersion curves give a criterion
for the crystal stability and indicate, through the prediction of soft
modes, the possible phase transitions. Indeed, if all phonon square
frequencies [$\omega^2(m,\mathbf{q})$] are positive, the crystal is
locally stable. However, if it appears that some
[$\omega^2(m,\mathbf{q})$] are imaginary (soft modes), then the system
is unstable. The phonon dispersion curves of Ni$_2$TiSn and NiTiSn,
are displayed in Fig. \ref{VBS} along several high-symmetry
directions. First, no soft mode is predicted by our calculations in
the whole Brillouin zone at ambient pressure and 0~K. This result
supports the absence of a temperature-driven displacive phase
transition for both compounds, in agreement with experiments. Then,
the acoustic branches of Ni$_2$TiSn and NiTiSn show a significant
dispersion. In the case of Ni$_2$TiSn, they exhibit a noticeable
mixing with the first low-frequency optical modes in the whole
Brillouin zone. So, this compound may have interesting thermal
expansion properties~\cite{PatAgCo}. Surprisingly, only the thermal
expansion of NiTiSn has been investigated~\cite{Wee}. The contribution of each kind of atom to each branch 
is also displayed in Fig. \ref{VBS} using a color code. We observe
that the dispersion curves of NiTiSn are dominated by Sn-atoms below
155~cm$^{-1}$ while both Ni and Ti-atoms are mainly involved above
this frequency. In Ni$_2$TiSn, a clear identification of its atomic
contributions is much more difficult due to the
significant dispersion of the branches below 200~cm$^{-1}$. This
dispersion leads to a quite continuous profile in its vibrational density-of-states
(VDOS) spectrum between 0 and 300~cm$^{-1}$ (see Fig. \ref{VDOS}). In contrast,
the optical phonon branches of NiTiSn are quite dispersionless within this
frequency range, leading to well-isolated peaks in its VDOS
spectrum. Indeed, NiTiSn shows a more discrete profile which can be
decomposed into three frequency ranges: (i) a peak centered at
100~cm$^{-1}$ with a shoulder at 125~cm$^{-1}$, (ii) a phonon gap
between 150 and 175 cm$^{-1}$, and (iii) a complex multipeak structure up to
300~cm$^{-1}$. Finally, Ni$_2$TiSn has a significant DOS at low
frequencies whereas a low density of modes is observed in NiTiSn up to
the first peak around 100~cm$^{-1}$. This last observation suggests
that the NiTiSn lattice should have the highest stiffness.

\begin{figure}
\begin{center}
\includegraphics[width=16cm]{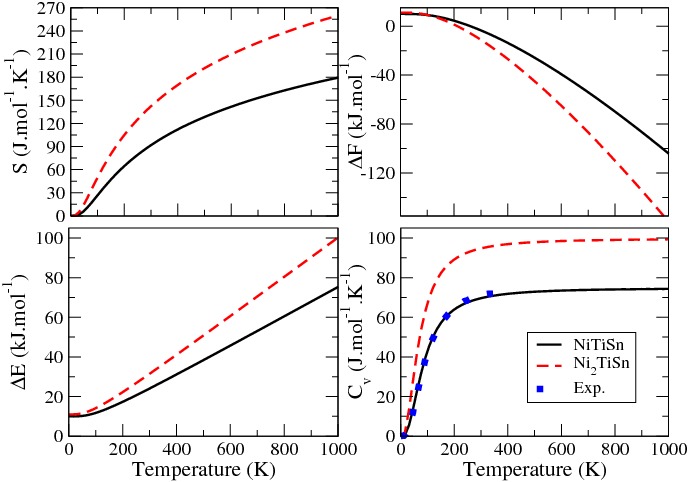}
\caption{GGA calculations of phonon contributions to entropy $S$, Helmholtz free energy $\Delta F$, internal energy $\Delta E$, and constant-volume specific heat $C_v$ of NiTiSn (solid black line) and Ni$_2$TiSn (dashed red line). Experimental data~\cite{these} (blue squares) of NiTiSn are also displayed for $C_v$.}
\label{Fig}
\end{center}
\end{figure}

From these DOS, the phonon contributions to entropy, Helmholtz free
energy, internal energy, as well as the constant-volume specific heat
can be derived according to the quasi-harmonic
approximation~\cite{Maradudin}. Phonon contributions to the entropy of
both compounds are displayed in Fig. \ref{Fig}. NiTiSn is found to have the
lowest entropy over the whole temperature range due to the higher
stiffness of its lattice. The zero-temperature value of the Helmholtz free
energy and the internal energy do not vanish due to the zero-point
motion. It can be calculated from the asymptotic equation:
\begin{equation}
 \Delta F_0 = \Delta E_0 = 3nN \int_0^{\omega_L} \frac{\hbar\omega}{2}g(\omega)d\omega,
\end{equation}
where ${\omega_L}$ is the largest phonon frequency, $n$ is the number
of atoms per unit cell, $N$ is the number of unit cells and
$g(\omega)$ is the VDOS. The latter is normalized according to:
$\int_0^{\omega_L} g(\omega)d\omega=1$. We find $\Delta F_0 = \Delta
E_0 = $11.02~kJ.mol$^{-1}$ for Ni$_2$TiSn and 9.98~kJ.mol$^{-1}$ for
NiTiSn. When increasing the temperature, the variation of the internal
energy shows that the overall profiles between both compounds are
similar: it increases with increasing temperature, the energy of
Ni$_2$TiSn being the highest. In contrast, the variation of their free
energy has an opposite trend. The Ni$_2$TiSn free energy is clearly
the lowest above 150~K due to its higher entropy contribution. The
constant-volume specific heats (C$_v$) are displayed in Fig. \ref{Fig} with
the available experimental data reported for NiTiSn~\cite{these}. For Ni$_2$TiSn,
only measurements of C$_v$ below 25~K have been reported in the
literature to our knowledge~\cite{aliev,Boff}. First, we observe that
our calculations reproduce with a very good agreement the experimental
specific heat of NiTiSn below room temperature. Above this
temperature, some discrepancies are expected as the anharmonic effects
(like thermal expansion) should be explicitly considered. At high
temperatures, the specific heats approach the classical Dulong and
Petit asymptotic limit: $C_v(T\rightarrow\infty)=
$99.77~J.mol$^{-1}$.K$^{-1}$ for Ni$_2$TiSn and
$C_v(T\rightarrow\infty)= $74.83~J.mol$^{-1}$.K$^{-1}$ for NiTiSn. The
Debye temperature has also been calculated using a linear fit of $C_v$
with respect to $T^3$ at very low temperatures ($T<$ 4~K). We expect
that NiTiSn has the highest Debye temperature since it is closely
related to the stiffness or the melting temperature of a material. Our
GGA calculated Debye temperatures are consistent with this assumption,
as we found: $\theta_D= $ 360~K (resp. $\theta_D= $332~K) for NiTiSn
(resp. Ni$_2$TiSn). These values are in fair agreement with the
experimental ones: $\theta_D=$ 417~K for NiTiSn~\cite{Kuentzler} and
$\theta_D=$ 290$\pm$6~K for Ni$_2$TiSn~\cite{Boff}. We have not
considered the Debye temperatures measured by Aliev {\it et
  al.}~\cite{aliev} for this comparison since Kuentzler {\it et
  al.}~\cite{Kuentzler} reported a possible error in their
calculations.

\subsection{Mechanical properties}

The knowledge of bulk, shear and Young's moduli and Poisson's ratio of
materials is desirable for possible applications, since strong
materials are preferred to weak ones. These mechanical properties are
usually derived from the elastic contants. Elastic constants can be
described by a fourth-rank tensor ($C$), relating the stress tensor
($\sigma$) to the strain tensor ($\eta$), via the generalized Hooke's
law:
\begin{equation}
C_{\alpha\beta}=\frac{\partial \sigma_\alpha}{\partial \eta_\beta},
\end{equation}
where $\alpha, \beta = $1,2,...,6 denote the Cartesian directions
given in Voigt notation. This equation can be splitted into two main
contributions such as:
\begin{equation}
\label{el}
C_{\alpha\beta}=\left. \frac{\partial \sigma_\alpha}{\partial \eta_\beta}\right|_u + \sum_\kappa \frac{\partial \sigma_{\alpha}}{\partial u_\alpha(\kappa)}\frac{\partial u_\alpha(\kappa)}{\partial \eta_\beta}.
\end{equation}
The first term is the frozen (clamped) ion elastic tensor, whereas the
second term includes contributions from force-response internal stress
and displacement-response internal strain. The second term of Eq. \ref{el}
also accounts for the ionic relaxations in response to strain
perturbations. The addition of the two contributions is the
relaxed-ion elastic tensor, $C$. In cubic space groups, these tensors
have only three independent elements: $C_{11}$, $C_{12}$ and $C_{44}$.
The effective elastic moduli of polycrystalline aggregates are usually
calculated by two approximations due to Voigt~\cite{voigt} (V) and
Reuss~\cite{reuss} (R) who respectively assume an uniform strain or
stress throughout the polycrystal. Hill~\cite{hill} has shown that the
Voigt and Reuss averages are limits and suggested that the actual
effective moduli can be approximated by the arithmetic mean of the two
bounds, referred to as the Voigt-Reuss-Hill (VRH) values
\cite{Wu}. The explicit expressions of the bulk ($B$) and shear ($G$)
moduli as a function of elastic constants for cubic systems can be
found everywhere~\cite{grimvall}. However for the sake of clarity in
the discussion that follows , we report them here:
\begin{equation}
B_{VRH} = \frac{1}{2}(B_{V}+B_{R}),\ G_{VRH} = \frac{1}{2}(G_{V}+G_{R}),
\end{equation}
where
\begin{equation}
B_{V,R} = \frac{1}{3}(C_{11}+2C_{12}),\ G_R = \frac{5(C_{11} - C_{12})C_{44}}{4C_{44} + 3(C_{11} - C_{12})},\ G_V =\frac{1}{5}(C_{11} - C_{12} + 3C_{44}).
\end{equation}

Elastic constants of Ni$_2$TiSn and NiTiSn, calculated at the LDA and
GGA levels, are listed in Table \ref{cij} with selected mechanical
properties such as: bulk, shear and Young's moduli, and Poisson's
ratio. In a recent publication, Roy {\it et al.}~\cite{vanderbilt}
have theoretically investigated the electro-mechanical coupling
coefficient of a large number of Heusler compounds. In particular,
they report the $C_{44}$ elastic constant of NiTiSn ($C_{44}= $62~GPa)
calculated using the same method of calculation (linear response) and
the same DFT code than ours. However, when comparing our results with
theirs, supposed to be obtained at the LDA level, we found a close
match with our GGA results, suggesting that their results have probably 
been obtained at the GGA level.

We observe that our LDA results are systematically larger than the GGA
ones. Thus, since the tendencies between the Heusler and half-Heusler
compounds are the same, we will focus our discussion on the GGA
results. Both compounds are mechanically stable because their elastic
constants satisfy the Born mechanical stability restrictions for cubic
structures~\cite{Born}: $C_{11} > |C_{12}|, C_{44} > 0, C_{11}
+ 2C_{12} > 0$. The relaxed-ion elastic constants (physical elastic
constants) are usually smaller than the clamped-ion ones, since the
additional internal relaxation allows some of the stress to be
relieved. However, this trend is only observed for C$_{44}$, both for
Ni$_2$TiSn and NiTiSn. Accordingly, their bulk modulus is independent
of any phonon contributions. Phonon contributions to the shear and
Young's moduli are the most important in the case of the half-Heusler
compound (see Table \ref{cij}).

In agreement with our analysis of the VDOS spectra reported in Sec.III.B,
NiTiSn has the highest Young's and shear moduli, suggesting that this
compound should be the stiffest. However, we observe that its bulk
modulus is also the smallest, which seems paradoxical. To explain this
{\it a priori} unusual behaviour, one has to focus on the crystal
structure of both compounds. The NiTiSn structure is
characterized by four interpenetrating fcc lattices one of which being 
unoccupied. Its Ti-Ni-Sn distances and its unit cell volume are smaller than the ones of Ni$_2$TiSn. 
Thus, the Ni$_2$TiSn structure is expected to be the most flexible. Accordingly,
Ni$_2$TiSn has the smallest Young's modulus, as it describes tensile elasticity.
For the same reasons, its shear modulus is also the smallest.
By contrast to Young's modulus, the bulk modulus describes volumetric
elasticity, or the tendency of an object to deform in all directions
when uniformly loaded in all directions. It can be considered as an
extension of Young's modulus to three dimensions. Thus, we expect that
the Ni-vacancies weaken the NiTiSn structure when the isotropic
pressure increases, leading Ni$_2$TiSn to have the largest bulk
modulus.

Poisson's ratios of the two compounds are in the range of the values
obtained for steel (NiTiSn) or magnesium (Ni$_2$TiSn). Globally the
Heusler compound is more incompressible than the half-Heusler, as
suggested by the value of their bulk modulus. Finally, brittle or
ductile behavior of both compounds has been estimated according to the
value of the $B/G$ ratio, as proposed by Pugh \cite{pugh}. If $B/G >$
1.75, a ductile behavior is predicted, otherwise the material behaves
in a brittle manner. We predict that both compounds are
ductile. Although Hichour {\it et al.} \cite{hichour} have reported,
using the stress-strain method and DFT calculations at the LDA level,
a brittle behaviour for NiTiSn ($B/G= $1.68), our result is, however,
consistent. Indeed, we cannot unambiguously conclude the brittle or
ductile behavior of NiTiSn using only the LDA results because the two sets of
calculations are very close to the limit fixed by Pugh considering the
accuracy of the methods used to compute the elastic constants (linear
response or stress-strain method, choice of XC--functional and 
pseudopotentials). Nevertheless, our GGA calculations
clearly support the ductile behavior of NiTiSn ($B/G= $2.03) and, as
expected, we found that the metallic compound (Ni$_{2}$TiSn) is more
ductile than the semiconducting one (NiTiSn). At the time being, the
lack of experimental data does not permit to conclude definitely on
the mechanical properties of these compounds.

\section{Conclusions}

We have performed a thorough comparison of the vibrational and
mechanical properties of the Heusler (Ni$_{2}$TiSn) and half-Heusler
compounds (NiTiSn) using density functional perturbation theory. The
calculation of the Raman and infrared spectra of NiTiSn allowed us to
clarify the debate reported in the literature on the assignment of its
modes. Based on the calculation of the phonon density-of-states, we
have demonstrated that the significant states of Ni$_2$TiSn at low
frequencies are at the origin of (i) its smaller free energy, (ii) its
higher entropy, and (iii) its lower Debye temperature, with respect to
NiTiSn. We also expect that Ni$_2$TiSn has a larger linear
thermal expansion than NiTiSn due to the more important mixing of its acoustic and
optic branches. We have also reported the mechanical
properties of the two compounds. In agreement with the analysis of the
VDOS spectra, NiTiSn has the highest Young's and shear moduli,
supporting that it should be the stiffest. Paradoxically, we
found that its bulk modulus is also the smallest. We suggest that
this unusual behaviour is related to the ordered Ni-vacancies that weaken the structure of the half-Heusler compound. Both for Ni$_2$TiSn and NiTiSn, we have evidenced that phonons only contribute to the C$_{44}$
elastic constant, indicating  that phonons affect their shear and Young's moduli, but do not contribute to their bulk modulus. In contradiction with recent LDA results~\cite{hichour} we find that both compounds show a ductile behavior, the largest being for the metallic one. 

In this article, we have provided some benchmark theoretical results. Going further in the
interpretation would require better experimental characterization of
Ni$_2$TiSn and NiTiSn compounds. We hope our work will
motivate such experimental studies on these industrially important
materials.

\section*{Acknowledgements}

This work has been realized with the support of the CINES and the HPC@LR 
computer centers in Montpellier. We thank Jean-Claude T\'edenac for helpful
discussions.



\clearpage
\begin{table}
\begin{center}
\begin{tabular} {lcccccccccccc}
\hline
\hline
&&\multicolumn{5}{c}{Experiments} && \multicolumn{5}{c}{Calculations}   \\
\cline{3-7}\cline{9-13}
&&\multicolumn{2}{c}{Raman} && \multicolumn{2}{c}{Infrared} && \multicolumn{2}{c}{GGA} && \multicolumn{2}{c}{LDA} \\
\cline{3-4}\cline{6-7}\cline{9-10}\cline{12-13}
   && Ref. \cite{Mestres} & Ref. \cite{Popovic} && Ref. \cite{Mestres} & Ref. \cite{Popovic} && Present  & Ref.~\cite{Wee}&& Present &Ref.~\cite{Wee}\\
   && (80 K) & (300 K) && (80 K) & (300 K) && (0 K) & (0 K) && (0 K) & (0 K)\\           
\hline
TO1 && 220   & 222   &&  -    & 222.8   && 205 & 219 &&  230   & 236\\
TO2 && 255   & 254   && 266 & 257     && 248 & 248  &&  269  & 260\\ 
LO1 && -     & -     &&   -    & 223.2  && 206 & 220      &&  231  & 237\\
LO2 && 285   & -     &&  292 & 287.5    && 284 & 281  &&  284 &293\\
\hline
\hline
\end{tabular}
\end{center}
\caption{Zone-center phonon frequencies (in cm$^{-1}$) of NiTiSn.}
\label{freq}
\end{table}

\begin{table}[htbp]
\begin{center}
\begin{tabular} {lcclcc}
\hline
\hline

     & $\omega_m$ &$a$ &   &  $\omega_m$ & $a$ \\
     & (cm$^{-1}$)   & (Bohr$^{-3/2}$) && (cm$^{-1}$)   & (Bohr$^{-3/2}$) \\
\hline
TO1  & 230 & 0.0087  & TO2  & 269 &   0.0207      \\
LO1  & 232 &  0.0115 & LO2  & 310 &   -0.0234  \\
\hline
\hline
\end{tabular}
\end{center}
\caption{Raman susceptibility ($a$) of NiTiSn phonon modes ($\omega_m$) calculated using LDA. }
\label{sus}
\end{table}

\begin{table}[htbp]
\begin{center}
\begin{tabular} {lccccccc}
\hline
\hline
&\multicolumn{3}{c}{GGA} && \multicolumn{3}{c}{LDA} \\
\cline{2-4}\cline{6-8}
     & $S_m$ & $Z_m^\ast$ & $\varepsilon^0$ &&  $S_m$ & $Z_m^\ast$ & $\varepsilon^0$\\
\hline
$\varepsilon^\infty$&&&    22.51   && &&   19.88 \\
\hline
TO1  &      1.21    & 1.24  & 0.47  && 2.29 & 1.74 & 0.83   \\
TO2  &     24.93  &  4.93 &6.72    &&  23.11 & 4.69 & 6.12\\
Total (Phonons)&      &  & 7.19     && && 6.95  \\
\hline
Total&         &        & 29.70          &&  && 26.83  \\
\hline
\hline
\end{tabular}
\end{center}
\caption{Phonon contributions to static dielectric constant ($\varepsilon^0$) of NiTiSn. Mode oscillator strengths ($S_m$, $\times$10$^{-5}$ $a.u.$) and mode effective charges~\cite{Hermet06} ($Z_m^\ast$) are also reported. Note: 1 $a.u. =$ 253.2638413 m$^3$.s$^{-2}$.}
\label{FOS}
\end{table}

\begin{table}
\begin{center}
\begin{tabular} {lcccccccccccc}
\hline
\hline
 &\multicolumn{5}{c}{Ni$_2$TiSn} && \multicolumn{6}{c}{NiTiSn}\\
\cline{2-6}\cline{8-13}
    & \multicolumn{2}{c}{GGA} && \multicolumn{2}{c}{LDA} &&  \multicolumn{2}{c}{GGA} && \multicolumn{3}{c}{LDA} \\
\cline{2-3}\cline{5-6}\cline{8-9}\cline{11-13}
                & Clamped & Relaxed &&  Clamped & Relaxed && Clamped & Relaxed && Clamped & \multicolumn{2}{c}{Relaxed} \\
\hline		
$C_{11}$  & 172.28 & 172.28 && 218.34 & 218.34 && 196.41 & 196.41 && 273.50 & 273.50 &(264.94)\\
$C_{12}$  &  127.41& 127.41 && 172.71 & 172.71 && 82.16 & 82.16     && 92.25 & 92.25 &(89.24)\\
$C_{44}$  &  75.27  & 75.07  &&  98.24  & 95.39   && 77.69  & 60.61    && 100.90 & 83.29 &(87.85)\\
$B_{V,R}=B$&142.37 & 142.37&& 187.92 & 187.92 && 120.24 & 120.24 && 152.67 & 152.67 &(147.81)\\
$G_V$   & 54.14 & 54.02 && 68.07 & 66.36 &&  69.46 & 59.22 && 96.79 & 86.22 &\\
$G_R$  & 38.76 & 38.73 && 42.30 & 41.98 && 67.91 & 59.17 && 96.52 & 86.08 &\\
$G$ & 46.45 & 46.37 && 55.19 & 54.17 &&68.69 & 59.19 && 96.66 & 86.15 &(87.84)\\
$E$ & 125.67 & 125.49 && 150.80 & 148.26 && 173.10 & 152.54 && 239.44 & 217.53 &(219.96)\\
$\nu$ & 0.35 & 0.35 && 0.37 & 0.37 && 0.26 & 0.29 && 0.24 & 0.26 &(0.25)\\
$B/G$ & 3.07 & 3.07 && 3.41 & 3.47 && 1.75 & 2.03 && 1.58 & 1.77 &(1.68)\\
\hline
\hline
\end{tabular}
\end{center}
\caption{Calculated clamped-ion and relaxed-ion elastic constants ($C$) and related mechanical properties: bulk ($B$), shear ($G$) and Young's ($E$) moduli and Poisson's ($\nu$) ratio. All these quantities are given in GPa except for $B/G$ and Poisson ratio which are dimensionless. Calculated LDA values from Hichour {\it et al.}~\cite{hichour} are in brackets. The subscripts $V$ and $R$ denote the Voigt and Reuss approximations.}
\label{cij}
\end{table}

\clearpage

\clearpage

\clearpage

\clearpage

\clearpage

\clearpage

\clearpage


\begin{thebibliography}{00}

\bibitem{pearson} P. Villars, L. D. Calvert, ``Handbook of Crystallographic Data for Intermetallic Phases'', ASM, Metals Park, OH, Release 2010/2011.

\bibitem{properties} T. Graf, C. Felser, S. S. P. Parkin, {\it Progress in Solid State Chemistry}, {\bf 2011}, {\it 39}, 1-50. 

\bibitem{Felser} C. Felser, G. H. Fecher, B. Balke, {\it Angew. Chem. Int Ed} {\bf 2007}, {\it 46}, 668-699. 

\bibitem{Colinet} C. Colinet, P. Jund, J. C. Tedenac, {\it to be published}. 

\bibitem{Xia}  Y. Xia, S. Bhattacharya, V. Ponnambalam, A. L. Pope, S. J. Poon, T. M. Tritt, {\em J. Appl. Phys.}, {\bf 2000}, {\it 88}, 1952

\bibitem{thermo} C. Uher, J. Yang, S. Hu, D. T. Morelli, G. P. Meisner, {\it Phys. Rev. B} {\bf 1999}, {\it 59}, 8615-21; Y. Kimura, H. Ueno, Y. Mishima, {\em J Electron Mater} {\bf 2009} {\it 38}, 934; W. Xie, Q. Jin, X. Tang, {\it J Appl Phys} {\bf 2008}, {\it 103}, 043711; M. Schwall, B. Balke, {\it Appl Phys Lett} {\bf 2011}, {\it 98}, 042106; K. Mastronardi, D. Young, C.-C. Wang, P. Khalifah, R. J. Cava, A. P. Ramirez, {\it Appl Phys Lett} {\bf 1999}, {\it 74}, 1415

\bibitem{Wee} D. Wee, B. Kozinsky, B. Pavan, M. Fornari, {\it J. Elect. Mat.} {\bf 2012}, {\it 41}, 977-983.

\bibitem{thermo2} Y. Xia, V. Ponnambalam, S. Bhattacharya, A. L. Pope, S. J. Poon, T. M. Tritt, {\it J Phys. Condens Matter} {\bf 2001}, {\it 13}, 77-89; S. Bhattacharya, A. L. Pope, R. T. Littleton, T. M. Tritt, V. Ponnambalam, Y. Xia, S. J. Poon, {\it Appl Phys Lett} {\bf 2000}, {\it 77}, 2476; P. Qui, X. Huang, X. Chen, L. Chen, {\em J Appl Phys} {\bf 2009}, {\em 106}, 103703.

\bibitem{Popovic} Z. V. Popovic, G. Kliche, R. Liu, F. G. Aliev, {\it Solid State Commun.} {\bf 1990}, {\it 74}, 829-832.

\bibitem{Mestres} N. Mestres, J. M. Calleja, F. G. Aliev, A. I. Belogorokhov, {\it Solid State Comm.} {\bf 1994}, {\it 91}, 779-784.

\bibitem{ABINIT} X. Gonze, B. Amadon, P. M. Anglade, J. M. Beuken, F. Bottin, P. Boulanger, F. Bruneval, D. Caliste, R. Caracas, M. Cote, {\it et al.} {\it Comput. Phys. Comm.} {\bf 2009}, {\it 180}, 2582-2615.

\bibitem{PBE} J. P. Perdew, K. Burke, M. Ernzerhof, {\it Phys. Rev. Lett.} {\bf 1996}, {\it 77}, 3865-3868.

\bibitem{PW} J. P. Perdew, Y. Wang, {\it Phys. Rev. B} {\bf 1992}, {\it 45}, 13244-13249.

\bibitem{TM} N. Troullier, J. L. Martins, {\it Phys. Rev. B} {\bf 1991}, {\it 43}, 1993-2006.

\bibitem{Monkhorst} H. J. Monkhorst, J. D. Pack, {\it Phys. Rev. B} {\bf 1976}, {\it 13}, 5188-5192.

\bibitem{Gonze97}  X. Gonze, C. Lee, \textit{Phys. Rev. B} \textbf{1997}, {\it 55}, 10355-10368.

\bibitem{Gonze} X. Gonze, J.-C. Charlier, D. C. Allan, M. P. Teter, {\it Phys. Rev. B} {\bf 1994}, {\it 50}, 13035-13038.

\bibitem{Veithen04} M. Veithen, X. Gonze, Ph. Ghosez, {\it Phys. Rev. Lett.} {\bf 2004}, {\it 93}, 187401.

\bibitem{Hermet06} P. Hermet, M. Veithen, Ph. Ghosez, {\it J. Phys.: Condens. Matter} {\bf 2007}, {\it 19}, 456202.

\bibitem{BTO} P. Hermet, M. Veithen, Ph. Ghosez, {\it J. Phys.: Condens. Matter} {\bf 2009}, {\it 21}, 215901.

\bibitem{Note} The computation of the Raman tensor is presently restricted to semiconducting materials and to the LDA level inside ABINIT.

\bibitem{gorlich} E. A. G\" orlich, K. Latka, A. Szytula, D. Wagner, R. Kmiec, K. Ruebenbauer, {\it Solid State Commun.} {\bf 1978}, {\it 25}, 661-663.

\bibitem{ouardi} S. Ouardi, G. H. Fecher, B. Balke, X. Kozina, G. Styganyuk, C. Felser, S. Lowitzer, D. K\" odderitzsch, H. Ebert, E. Ikenaga, {\em Phys. Rev. B} {\bf 2010}, {\it 82}, 085108.

\bibitem{aliev} F. G. Aliev, V. V. Kozyrkov, V. V. Moshchalkov, R. V. Scolozdra, K. Durczewski, {\it Z. Phys. B--Condensed Matter} {\bf 1990}, {\it 80}, 353-357.

\bibitem{gap} V. A. Romaka, P. Rogl, V. V. Romaka, E. K. Hlil, Yu, V. Stadnyk, S. M. Budgerak, {\em Semiconductors} {\bf 2011}, {\em 45}, 850-856; L. L. Wang, L. Miao, Z. Y. Wang, W. Wei, R. Xiong, H. J. Liu, J. Shi, X. F. Tang, {\em J. Appl. Phys.} {\bf 2009}, {\em 105}, 013709; H. Hazama, R. Asahi M. Matsubara, T. Takeuchi,  {\em J. Electron. Mater.} {\bf 2010}, {\em 39}, 1549-1553; M. Ameri, A. Touia, R. Khenata, Y. Al-Douri, H. Baltache, {\em Optik} {\bf 2013}, {\em 124}, 570-574.

\bibitem{PatBFO} M. Goffinet, P. Hermet, D. I. Bilc, Ph. Ghosez, {\it Phys. Rev. B} {\bf 2009}, {\it 79}, 014403.

\bibitem{PatNiOH} P. Hermet, L. Gourrier, J.-L. Bantignies, D. Ravot, T. Michel, S. Deabate, P. Boulet, F. Henn, {\it Phys. Rev. B} {\bf 2011}, {\it 84}, 235211. 

\bibitem{PatAgCo} P. Hermet, J. Catafesta, J.-L. Bantignies, C. Levelut, D. Maurin, A. B. Cairns, A. L. Goodwin, J. Haines, \textit{J. Phys. Chem. C} \textbf{2013}, {\it 117}, 12848-12857.

\bibitem{Maradudin} A. A. Maradudin, E. W. Montroll, E. H. Weiss, I. P. Iaptova, {\it Theory of Lattice Dynamics in the Harmonic Approximation}, 2rd Eds., Academic, New York, 1971.

\bibitem{these} B. Zhong, {\it Master's Thesis}, Iowa State University, 1997. 

\bibitem{Boff} M. A. S. Boff, G. L. F.  Fraga, D. E.  Brandao, A. A. Gomes, T. A. Grandi, {\it Phys. Stat. Sol. (a)} {\bf 1996}, {\em 154}, 549.

\bibitem{Kuentzler} R. Kuentzler, R. Clad, G. Schmerber, Y. Dossmann, {\it J. Magn. Magn. Mater.} {\bf 1992}, {\it 104-107}, 1976-1978.

\bibitem{voigt} W. Voigt, Lehrbuch de Kristallphysik, Terubner, Leipzig, (1928).

\bibitem{reuss} A. Reuss, {\em Z. Angew. Math. Mech.} {\bf 1929}, {\it 9}, 49.

\bibitem{hill} R. Hill, {\em Proc. Phys. Soc. London A} {\bf 1952}, {\it 65}, 349.

\bibitem{Wu} Z. J. Wu, E. J. Zhao, H. P. Xiang, X. F. Hao, X. J. Liu, J. Meng, {\it Phys. Rev. B} {\bf 2007}, {\em 76}, 054115.

\bibitem{grimvall} G. Grimvall, "Thermophysical properties of materials'' Elsevier ed. (1999).

\bibitem{vanderbilt} A. Roy, J. W. Bennett, K. M. Rabe, D. Vanderbilt, {\em Phys. Rev. Lett.} {\bf 2012}, {\em 109}, 037602.

\bibitem{Born} M. Born and K. Huang, in Dynamical Theory of Crystal Lattices, Oxford U.P. (1955).

\bibitem{pugh} S. F. Pugh, {\em Philos. Mag.} {\bf 1954}, {\em 45}, 823.

\bibitem{hichour} M. Hichour, D. Rached, R. Khenata, M. Rabah, M. Merabet, A. H. Reshak, S. Bin Omran, R. Ahmed, {\em J. Phys. Chem. Solids} {\bf 2012}, {\em 73}, 975-981. 

\end{thebibliography}
\end{document}